\documentclass{iau_FM}
\usepackage{graphicx}
\usepackage{natbib}

\title[Massive Stars in the Far and Extreme Ultraviolet] 
{Massive Stars in the Far and Extreme Ultraviolet}

\author[A.A.C. Sander]   
{Andreas A.C. Sander$^1$}

\affiliation{$^1$Zentrum für Astronomie der Universität Heidelberg, Astronomisches Rechen-Institut, Mönchhofstr. 12-14, 69120 Heidelberg, Germany \\ email: {\tt andreas.sander@uni-heidelberg.de}} 

\pubyear{2022}
\jname{Astronomy in Focus, Focus Meeting 4} 
\editors{José Espinosa, ed.}
\begin{document}

\maketitle

\begin{abstract}
From the main sequence to their late evolutionary stages, massive stars spend most of their life as hot stars. Due to their high effective temperatures, the maximum of their emitted flux falls into the regime of ultraviolet (UV) wavelengths. Consequently, these stars emit a significant number of photons with energies sufficiently high enough to ionize hydrogen and potentially also other elements. As simple as these fundamental considerations are, as complex is a realistic estimate of the resulting ionizing fluxes, in particular for energies above 54 eV.  

Estimating the ionizing flux budget of hot stars requires accurate models of their spectral energy distributions (SEDs), covering in particular the far and extreme UV region. Modern atmosphere models that incorporate the so-called line-blanketing effect, i.e. taking into account the millions of lines from iron and other elements, yield a complex picture, illustrating that the SED of a hot, massive star often deviates significantly from a blackbody. The ubiquitous presence of stellar winds complicates the picture: the absorption of photons driving the mass outflow leads to flux being shifted to longer wavelengths, strongly affecting the flux budget at the highest energies.
On top of all these challenges, models estimating the ionizing fluxes of a whole population face the challenge of approximating massive star formation and evolution, which contain major unsolved puzzles often interwoven with open questions on the stellar scale. 

\end{abstract}

\firstsection 
\section{Hot, massive stars and their ionizing fluxes}

Massive stars spend most of their lifetime as hot stars with $T_\mathrm{eff} > 10\,$kK. Consequently, these stars have their flux maximum at ultraviolet (UV) wavelengths, providing an efficient source of flux for driving a stellar wind and ionizing their surrounding environment. Photometric and spectroscopic observations in the UV are particularly crucial to determine the parameters of hot, massive stars and their outflows. As the extreme ultraviolet (EUV) is largely inaccessible due to the presence of interstellar hydrogen, the far UV regime between 900 and 2000\,\AA\ contains the most viable diagnostics to photometrically constrain $T_\mathrm{eff}$ or infer the wind properties and ionizing fluxes if spectroscopy is available. 

The ionizing fluxes of hot stars are commonly quantified by integrating the photon flux beyond a frequency $\nu_\mathrm{edge}$, namely
\begin{equation}
  \label{eq:qdef}
   Q_\mathrm{edge} := \int\limits_{\nu_\mathrm{edge}}^\infty \frac{F_\nu}{h \nu} \mathrm{d}\nu
\end{equation}
with typically $\log Q_\mathrm{edge}$ given in the literature, taking the logarithm of the value of photons per second. Commonly used are $\log Q_0 \equiv \log Q_\mathrm{H\,\textsc{i}}$ for the hydrogen-ionizing flux beyond the Lyman edge, and $\log Q_1 \equiv \log Q_\mathrm{He\,\textsc{i}}$ as well as $\log Q_2 \equiv \log Q_\mathrm{He\,\textsc{ii}}$ for the ionizing fluxes beyond the edges of He\,\textsc{i} and He\,\textsc{ii} ionization. 

\section{Measuring Ionizing Fluxes}

Given the EUV inaccessibility, ionizing fluxes are mostly obtained by either measuring nebula emission lines or integrating the spectral energy distribution (SED) from an atmosphere model assigned to the source star(s) \citep[e.g.,][]{Ramachandran+2018}.

\begin{figure}[htb]
\begin{center}
   \includegraphics[angle=0,width=0.49\textwidth]{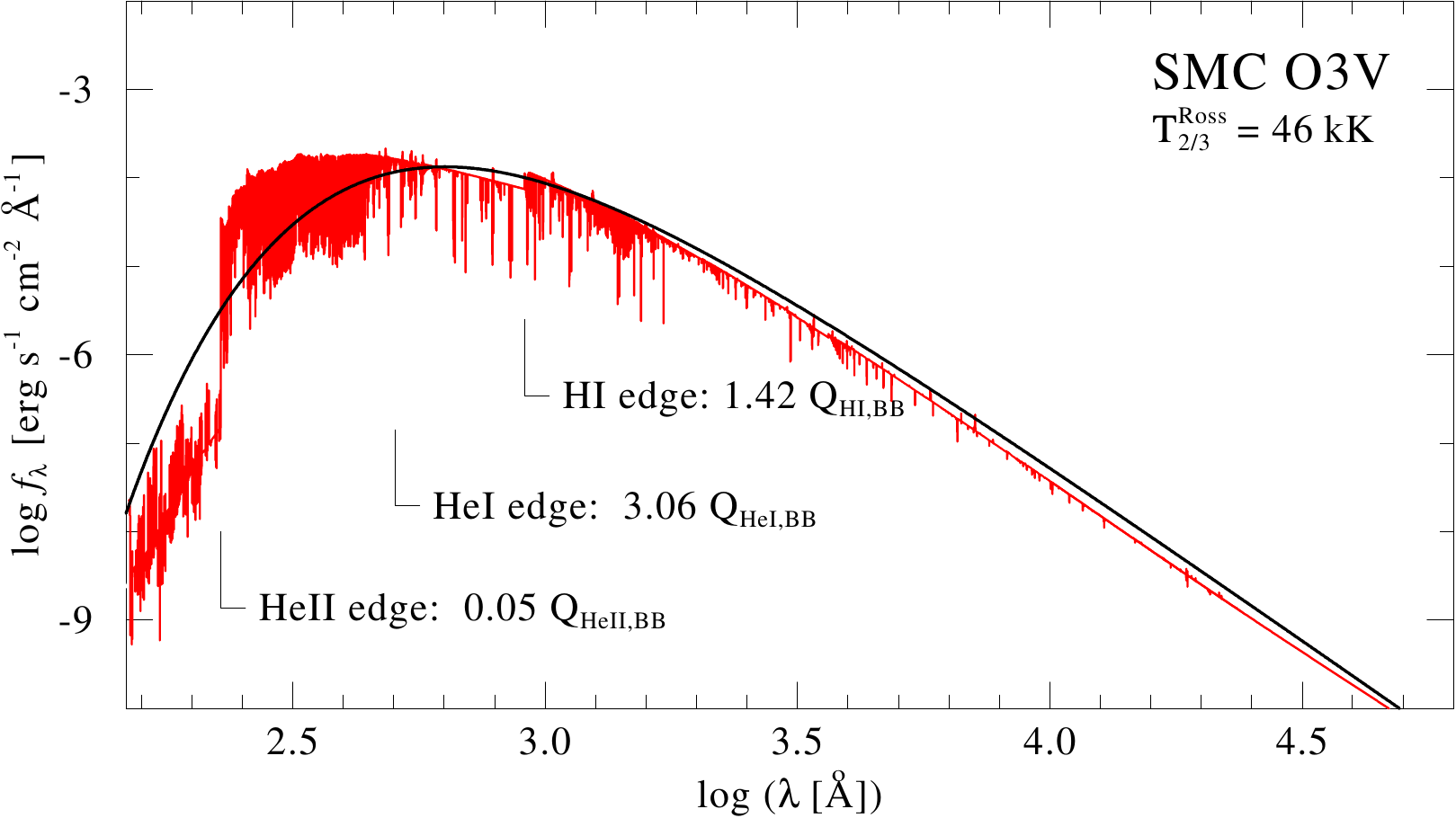}
	 \hfill
   \includegraphics[angle=0,width=0.49\textwidth]{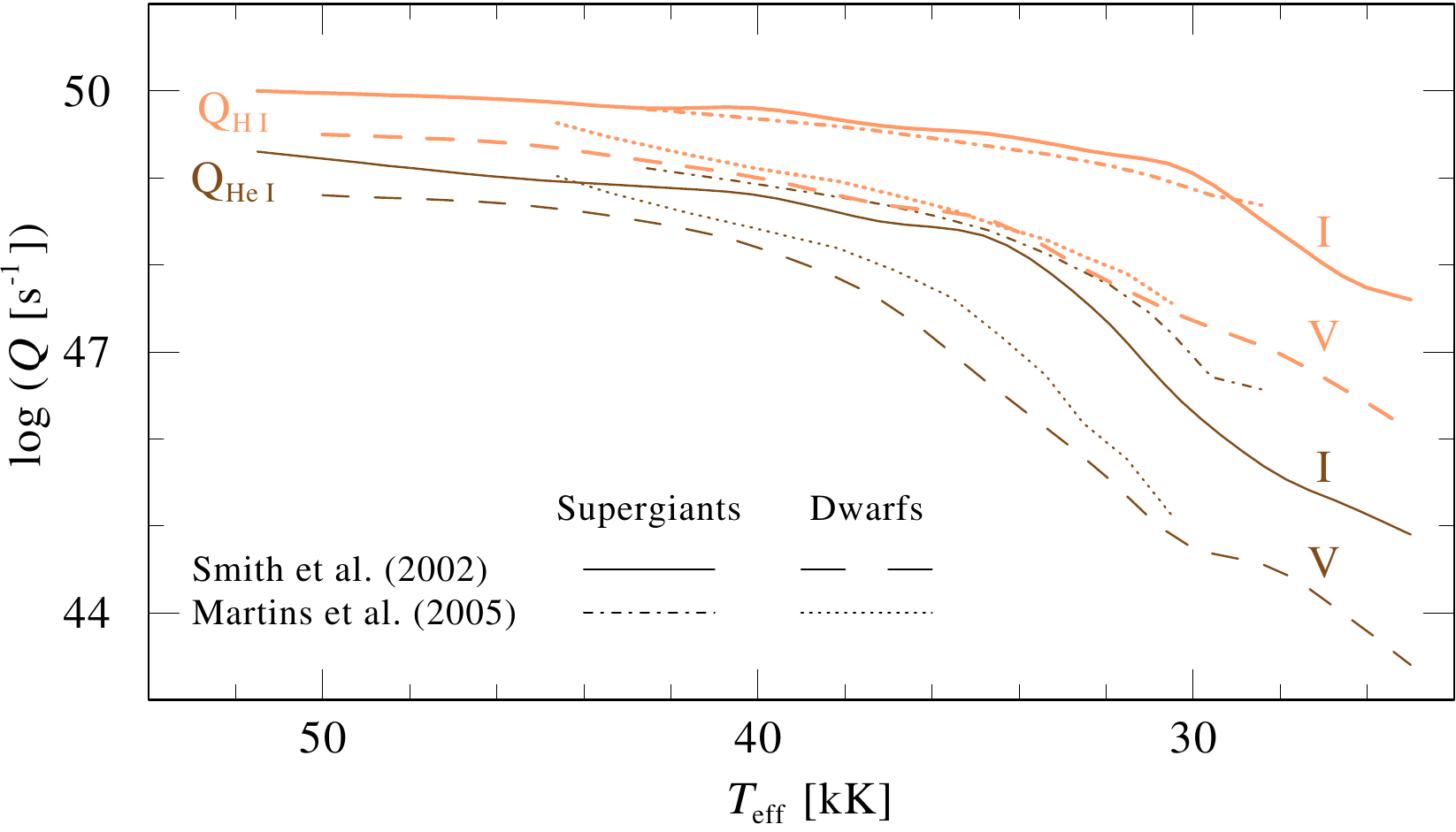}
     \caption{Left: Comparison of the spectral energy distribution (SED) from an atmosphere model (red) for an O3 SMC dwarf to a blackbody (black) of the same $T_\mathrm{eff}$. Right: Comparison of $Q_\mathrm{H\,\textsc{i}}$ and $Q_\mathrm{He\,\textsc{i}}$ for OB star models from \citet{Smith+2002} and \citet{Martins+2005}.}
   \label{fig:ionflux-intro}
\end{center}
\end{figure}

The use of atmosphere models is crucial as estimates from blackbodies can be severely off (cf.\ Fig.\,\ref{fig:ionflux-intro}). For $Q_\mathrm{H\,\textsc{i}}$ and $Q_\mathrm{He\,\textsc{i}}$, spectral subtype calibrations vary a bit among the literature (cf.\ right panel Fig.\,\ref{fig:ionflux-intro}) with many uncertainties existing for lower metallicities.

\section{Interplay between stellar winds and He\,II ionizing flux}

In contrast to $Q_\mathrm{H\,\textsc{i}}$ and $Q_\mathrm{He\,\textsc{i}}$, the He\,\textsc{ii} ionizing flux $Q_\mathrm{He\,\textsc{ii}}$ depends strongly on the wind strength of the stars and thus also on the metallicity $Z$. In strong, dense winds, He\,\textsc{iii} recombines, making the atmosphere opaque for $Q_\mathrm{He\,\textsc{ii}}$ photons \citep[e.g.][]{Schmutz+1992}. The absorbed flux also drives the wind and is re-emitted only at longer wavelengths. 

\begin{figure}[htb]
\begin{center}
   \includegraphics[angle=0,width=0.49\textwidth]{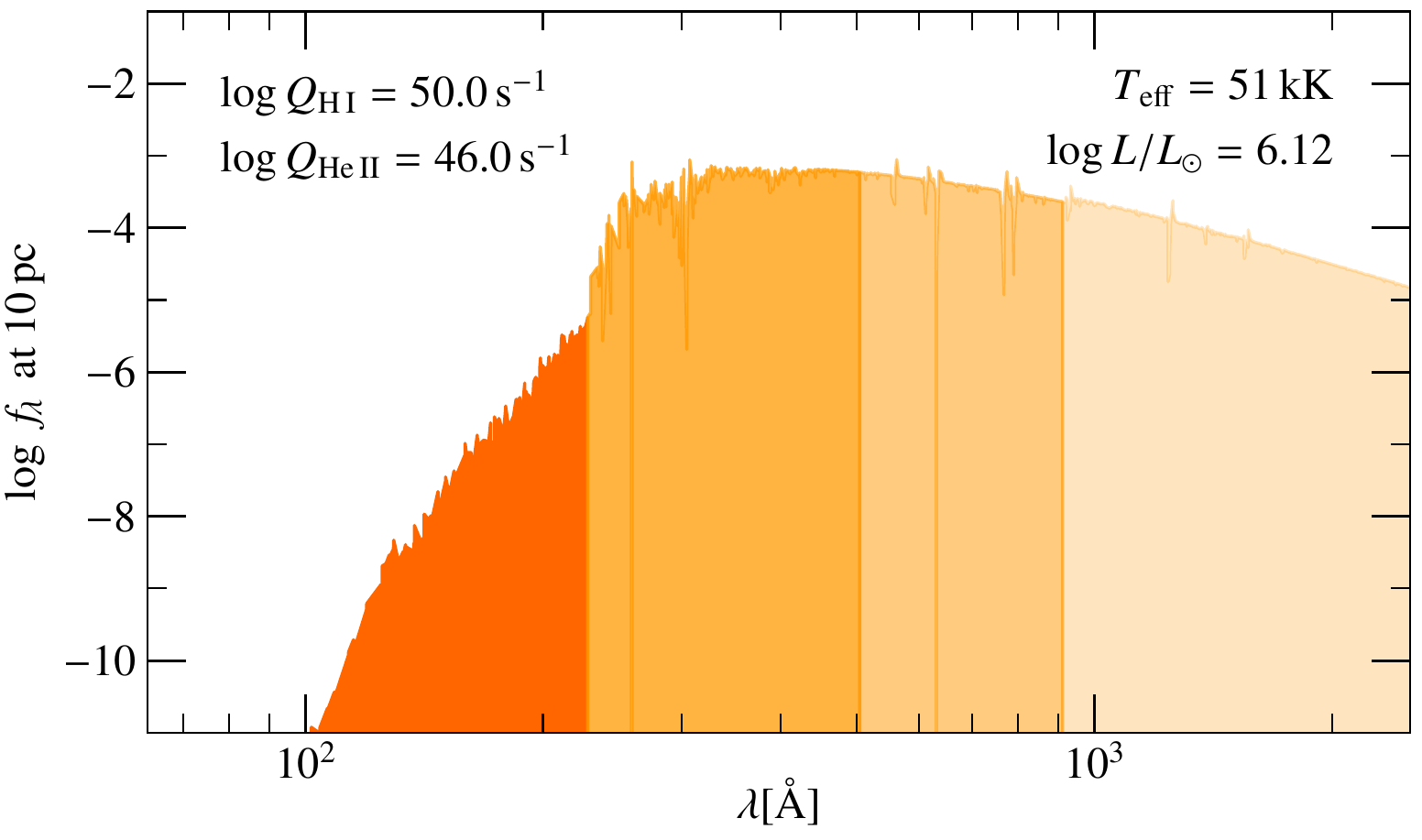}
	 \hfill
   \includegraphics[angle=0,width=0.49\textwidth]{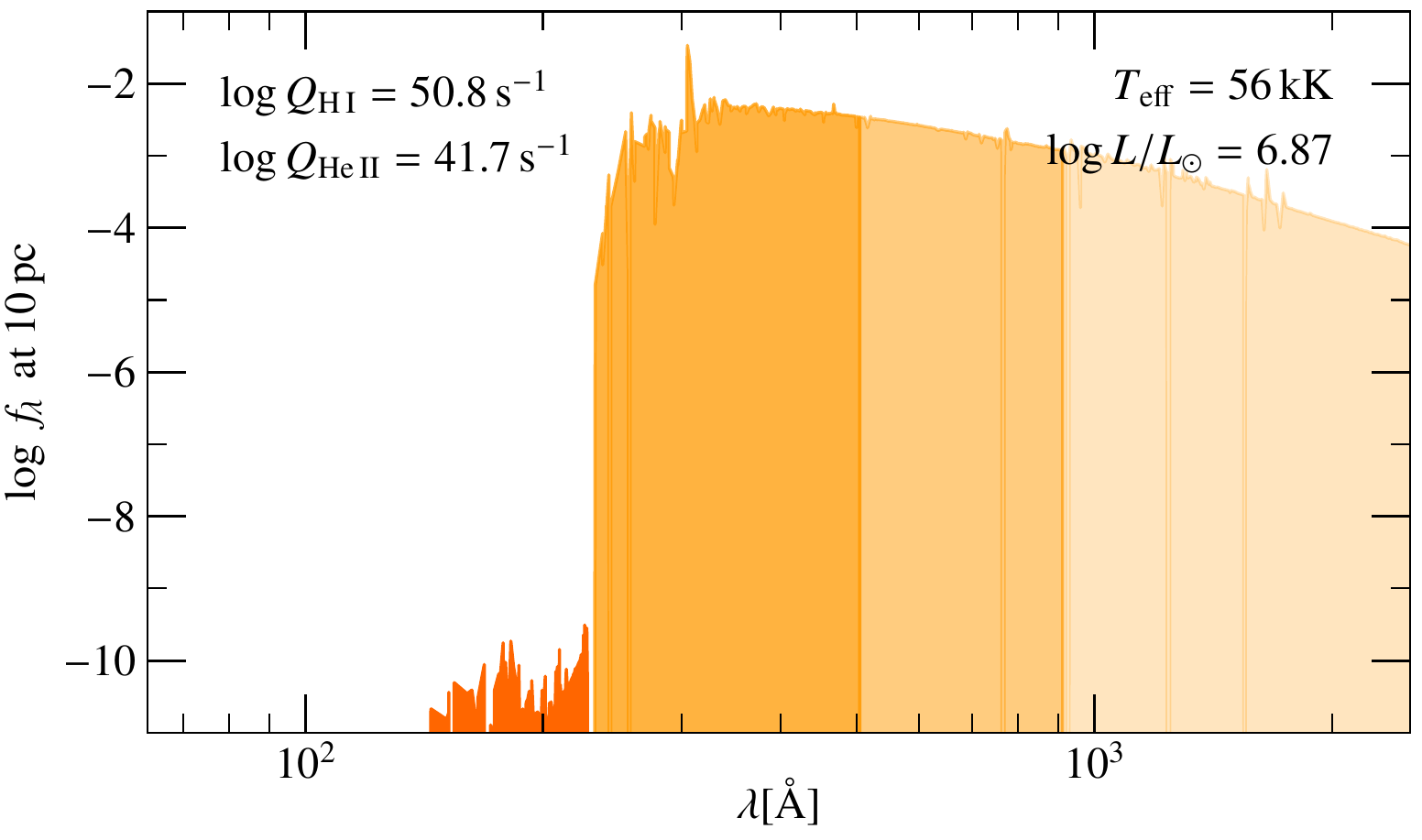}
     \caption{SED highlighting the flux beyond the ionization edges of H\,\textsc{i}, He\,\textsc{i}, and He\,\textsc{ii} for the O2 giant SSN9 in the SMC \citep[left panel, model from][]{Rickard+2022} and the WN5h star R136a1 \citep[right panel, model based on][]{Hainich+2014}.}
   \label{fig:ionflux-SSN9-R136a1}
\end{center}
\end{figure}

Apart from a high intrinsic temperature, stellar sources for $Q_\mathrm{He\,\textsc{ii}}$ thus require winds that are mainly optically thin. Consequently, most Wolf-Rayet (WR) stars are not very efficient sources of $Q_\mathrm{He\,\textsc{ii}}$, regardless of whether they are classical, helium-burning WRs or very massive stars (VMS) that might still be hydrogen-burning. This is illustrated in Fig.\,\ref{fig:ionflux-SSN9-R136a1}, where an O2 giant in the SMC provides orders of magnitude more He\,\textsc{ii} ionizing flux than the hotter and five times more luminous WN5h-star R136a1 in the LMC. 

\begin{figure}[htb]
\begin{center}
   \includegraphics[angle=0,width=0.49\textwidth]{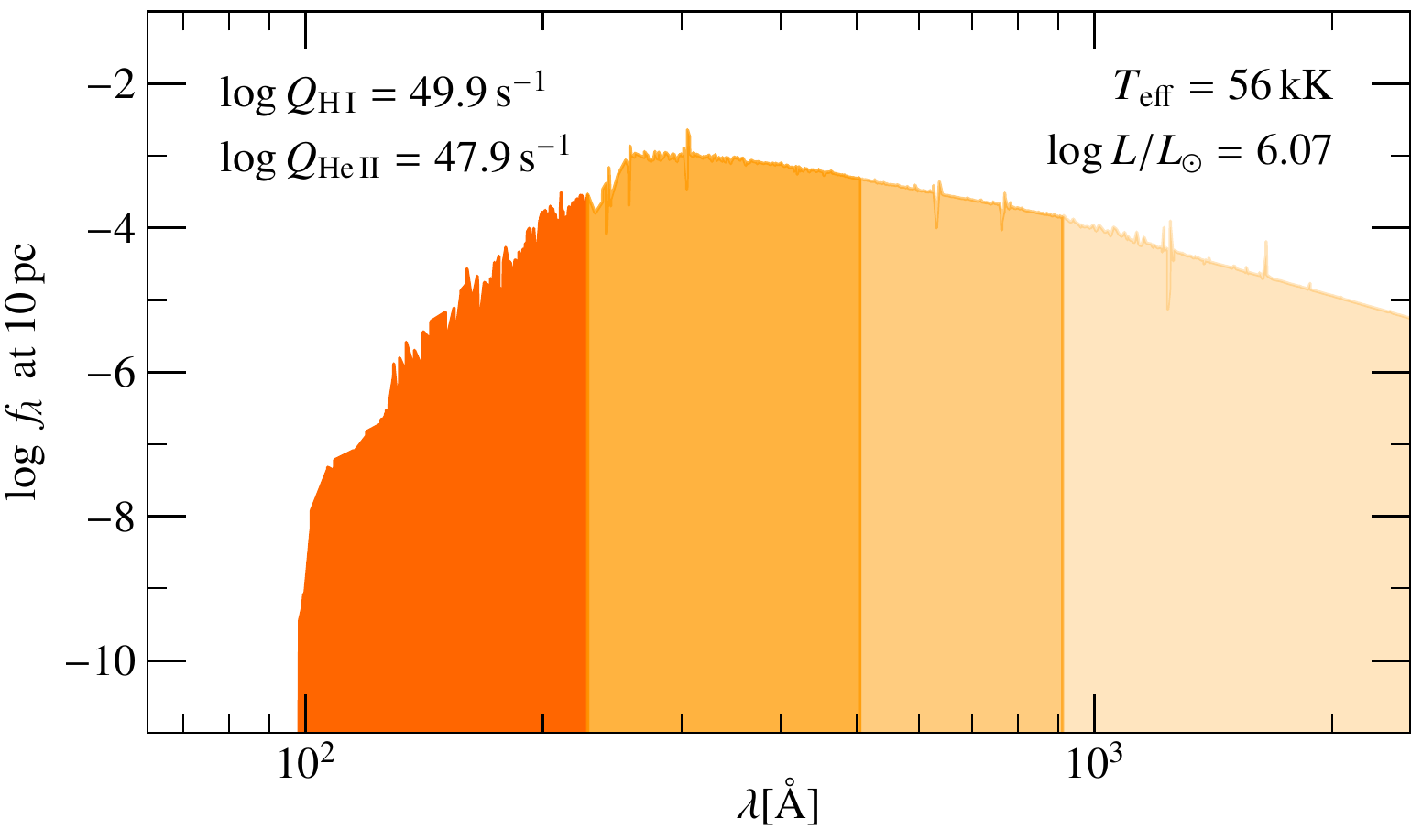}
	 \hfill
   \includegraphics[angle=0,width=0.49\textwidth]{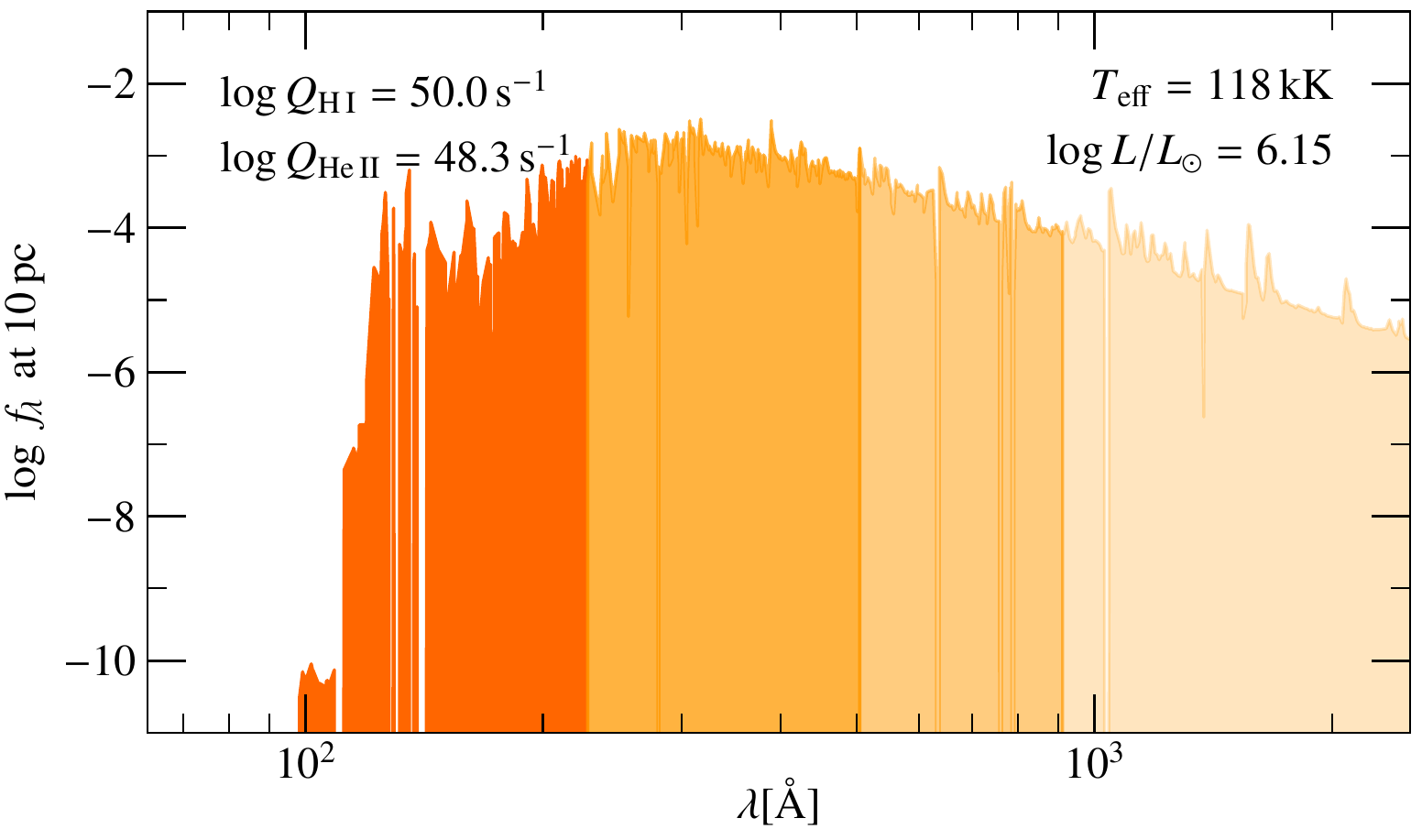}
     \caption{Same as Fig.\,\ref{fig:ionflux-SSN9-R136a1}, but for SMC AB\,1 \citep[WN3ha, left panel, model based on][]{Hainich+2015} and the WO4-star in SMC AB\,8 \citep[right panel, model based on][]{Shenar+2016}.}
   \label{fig:ionflux-AB1-AB8}
\end{center}
\end{figure}

\begin{figure}[htb]
\begin{center}
   \includegraphics[angle=0,width=0.49\textwidth]{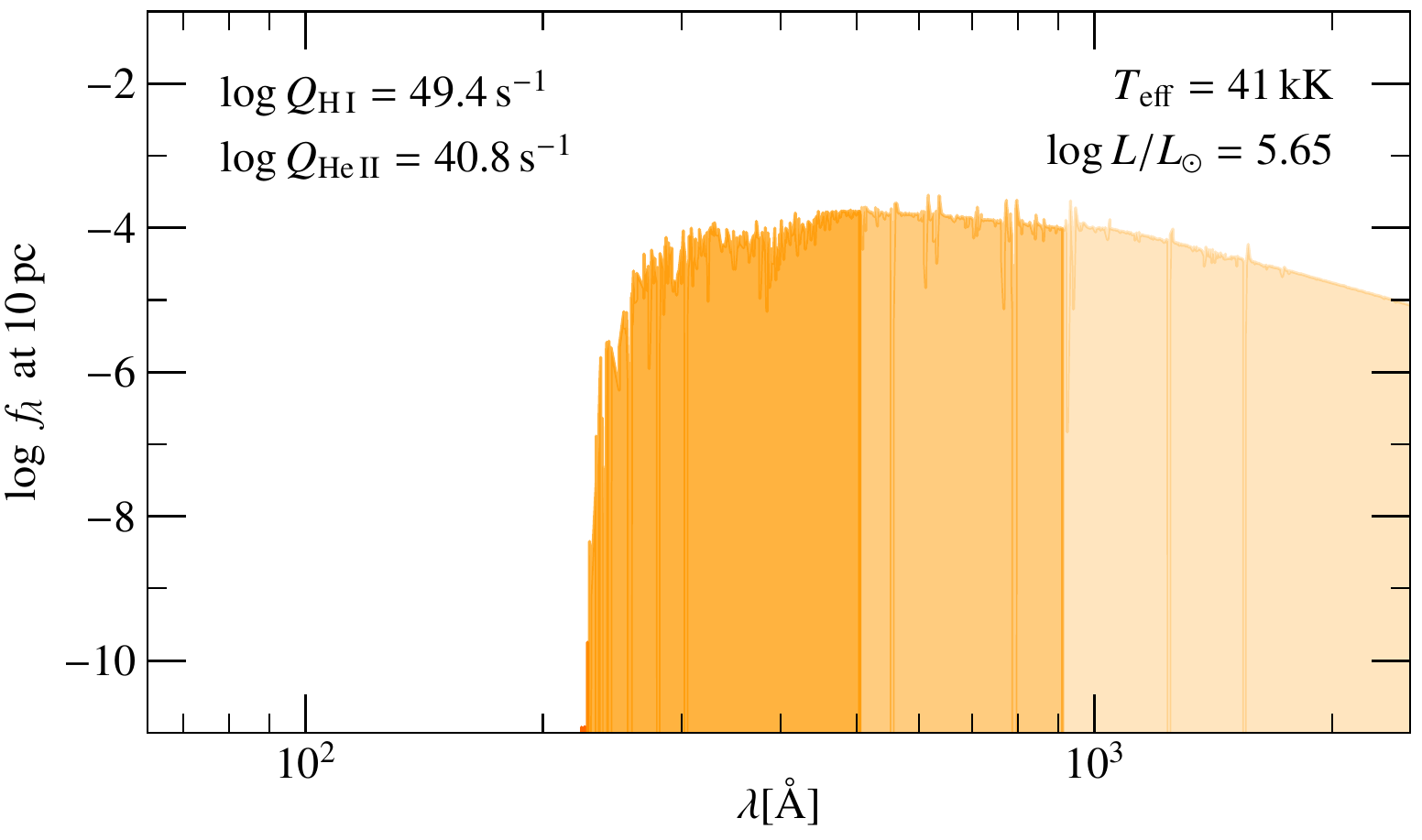}
	 \hfill
   \includegraphics[angle=0,width=0.49\textwidth]{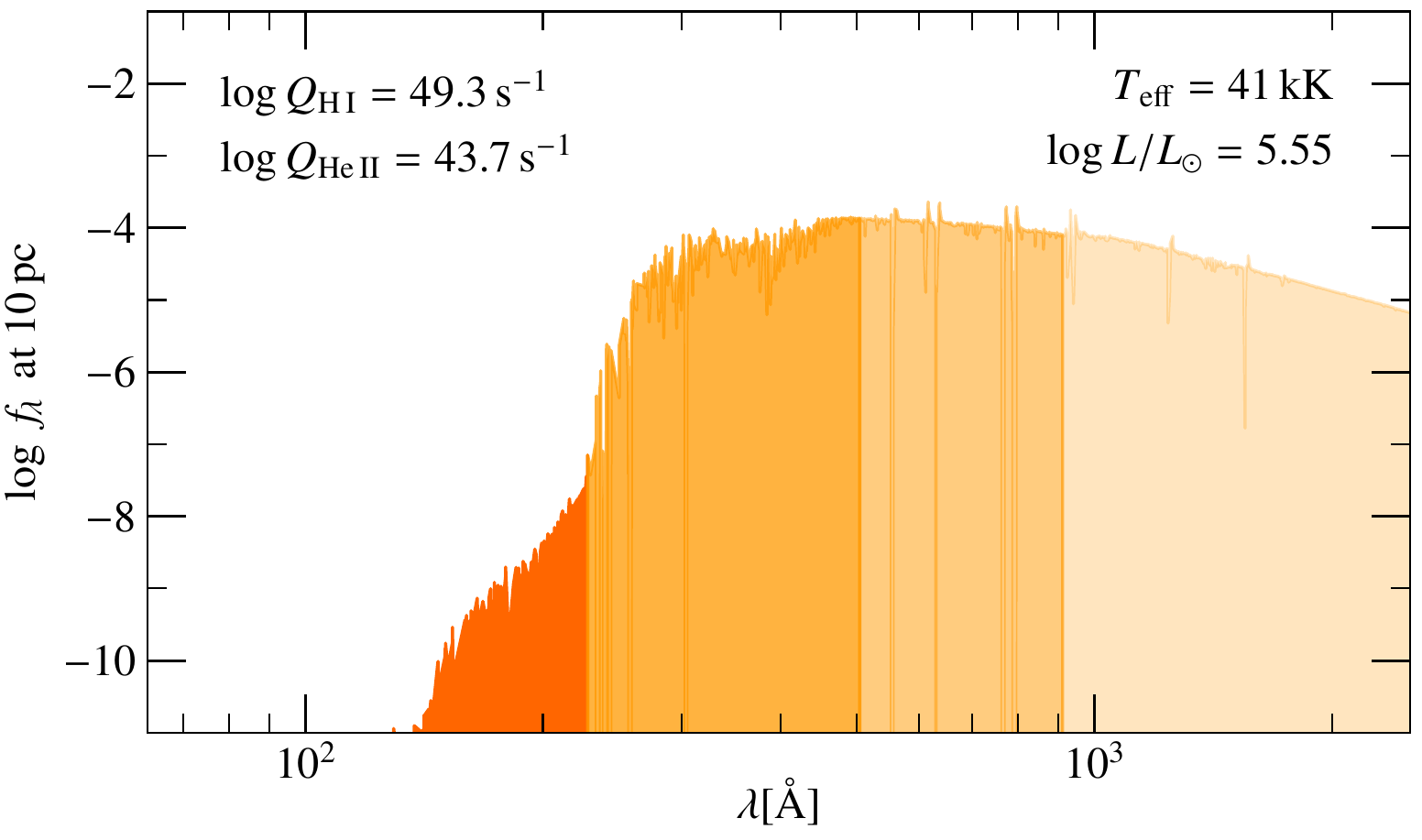}
     \caption{SEDs for the observed, partially stripped O4 primary in the SMC AzV 476 system (left panel) and the corresponding prediction based on a binary evolution model (right panel). The underlying models for the SED plots are taken from \citet{Pauli+2022}.}
   \label{fig:ionflux-AzV476}
\end{center}
\end{figure}

Nonetheless, some WR stars actually contribute huge amounts of $Q_\mathrm{He\,\textsc{ii}}$, such as weak-winded, early-type WNs and WO stars (see Fig.\,\ref{fig:ionflux-AB1-AB8} for examples). In all cases, the wind density is key here, making predictions difficult, not least due to the uncertainties in $\dot{M}$. An example involving binary evolution is shown in Fig.\,\ref{fig:ionflux-AzV476}, where the empirically obtained wind by \citet{Pauli+2022} is stronger than predicted by the evolutionary scenario for the system, leading to a reduction of $Q_\mathrm{He\,\textsc{ii}}$ by three orders of magnitude. 

In summary, stellar contributors of He\,\textsc{ii} ionizing flux require high temperatures and relatively thin winds. Known strong contributors are the most massive main sequence stars at low $Z$ as well as classical WR stars with weak winds (e.g. WN3ha, WO). A population of hot, hydrogen-depleted stars below the WR regime \citep[``stripped stars'', see e.g.][]{Goetberg+2020} could be important as well, but yet lacks observational confirmation.







\def\apj{{ApJ}}    
\def\nat{{Nature}}    
\def\jgr{{JGR}}    
\def\apjl{{ApJ Letters}}    
\def\aap{{A\&A}}   
\def\aaps{{A\&A Supplement}}   
\def\mnras{{MNRAS}}
\def\aj{{AJ}}
\def\pasa{{PASA}}
\def\pasp{{PASP}}
\def\ssr{{Space Science Reviews}}
\let\mnrasl=\mnras

\end{document}